\documentclass[structabstract]{aa} 
\usepackage{CJK}
\usepackage{amsmath}
\usepackage{graphicx}
\usepackage{txfonts}
\usepackage{natbib}
\usepackage{longtable}
\usepackage{lscape}
\usepackage{multirow}
\usepackage{pdfpages}
\usepackage{ulem}
\usepackage{mathtools}
\usepackage[colorlinks=true, allcolors=blue]{hyperref}

\bibpunct{(}{)}{;}{a}{}{,}

\newcommand{\kms}{km\,s$^{-1}$}

\newcommand{\degree}{$^{\circ}$}

\begin{document}
\title{First detection of CF$^{+}$ in the Large Magellanic Cloud}
\author{Yan Gong\inst{1,2}, Karl M. Menten\inst{1}, Arshia M. Jacob\inst{3,1}, Christian Henkel\inst{1}, C.-H.~Rosie Chen\inst{1}}

\institute{
Max-Planck-Institut f{\"u}r Radioastronomie, Auf dem H{\"u}gel 69, D-53121 Bonn, Germany \\
e-mail: [ygong; kmenten; chenkel; rchen]@mpifr-bonn.mpg.de
\and 
Purple Mountain Observatory, and Key Laboratory of Radio Astronomy, Chinese Academy of Sciences, 10 Yuanhua Road, Nanjing 210023, China e-mail: ygong@pmo.ac.cn
\and 
William H. Miller III Department of Physics \& Astronomy, Johns Hopkins University, 3400 North Charles Street, Baltimore, MD 21218, USA e-mail: ajacob51@jhu.edu
}

\date{Received date ; accepted date}

\abstract
{CF$^{+}$ has been established as a valuable diagnostic tool for investigating photo-dissociation regions (PDRs) and fluorine abundances in the Milky Way. However, its role in extragalactic environments remains largely uncharted.}
{Our objective is to explore the significance of CF$^{+}$ in the Large Magellanic Cloud (LMC) and assess its utility as a valuable probe for examining C$^{+}$ and fluorine abundances in external galaxies.}
{We performed pointed CF$^{+}$ observations toward an active star-forming region, N113 in the LMC, using the Atacama Pathfinder EXperiment 12~m sub-millimeter telescope.}
{We report the first discovery of CF$^{+}$ in the LMC through the successful detection of the CF$^{+}$ (2$\to$1) and (3$\to$2) lines. The excitation models indicate that CF$^{+}$ emission originates from dense PDRs characterized by an H$_{2}$ number density of $(0.5-7.9)\times 10^{4}$~cm$^{-3}$ in N113. Our observations provide the first constraint on the fluorine abundance in molecular clouds in the LMC, disclosing a value of $\lesssim 1.7\times 10^{-9}$. This value is about an order of magnitude lower than those previously measured toward red giants in the LMC, indicative of fluorine deficiency in the molecular gas. The estimated column density ratio between C$^{+}$ and CF$^{+}$ appears to be lower than the anticipated equilibrium ratio derived from the fluorine abundance in red giants. Both phenomena can be explained by the deficiency of CF$^{+}$ caused by the freeze-out of its primary chemical precursor, HF, onto dust grains.}
{The deficiency of CF$^{+}$ within molecular clouds suggests that the measurements presented in this work serve exclusively as conservative estimates, establishing lower bounds for both the fluorine abundance and C$^{+}$ column densities in external galaxies.}

\keywords{radio lines: ISM --- ISM: individual object (N113) --- ISM: molecules -- galaxies: ISM --- galaxies: Magellanic Clouds}

\titlerunning{First detection of CF$^{+}$ in the Large Magellanic Cloud}

\authorrunning{Y. Gong et al.}

\maketitle


\section{Introduction}
The abundances of both ionized carbon, C$^{+}$, and of fluorine are important parameters for understanding the life cycle of matter in galaxies. The far-infrared fine-structure line of C$^{+}$ at $\sim$158~$\mu$m (hereafter, [CII]) stands as one of the most crucial coolants in the Universe \citep[e.g.,][]{2003ApJ...587..278W,2011piim.book.....D}, and serves as a powerful tracer of star formation rates \citep{2015ApJ...800....1H} and CO-dark gas \citep[e.g.,][]{2010ApJ...716.1191W,2013A&A...554A.103P,2014A&A...561A.122L}, and photo-dissociation regions \citep[PDRs;][]{1999RvMP...71..173H,2022ARA&A..60..247W}. In parallel, the fluorine abundance can yield valuable insights into stellar nucleosynthesis history and chemical enrichment of a region \citep[e.g.,][]{2003AJ....126.1305C,2021NatAs...5.1240F}. 
Assessments of the fluorine abundance within molecular clouds have predominantly relied on the analysis of infrared and submillimeter spectra of the hydrogen fluoride molecule HF  \citep[e.g.,][]{1997ApJ...488L.141N,2010A&A...518L.108N}. Because HF is the only neutral diatomic hydride that can be produced by an exothermic reaction with H$_{2}$ \citep[e.g.,][]{2009ApJ...706.1594N,2016ARA&A..54..181G}, HF data can be used as a proxy to determine H$_{2}$ column densities of CO-dark molecular gas \citep[e.g.,][]{2019A&A...631A.117K}. However, the rest frequencies of the [CII] and HF rotational transitions are presently inaccessible for observations due to absorption by the Earth's atmosphere, since facilities such as \textit{Herschel} and \textit{SOFIA}, which were previously employed for investigating lines from both species, are no longer operational. Consequently, there arises a need for alternative tracers that can be readily observed using ground-based telescopes, thus affording additional insights into C$^{+}$ and fluorine abundances within the interstellar medium (ISM).

CF$^{+}$ is the second most abundant fluorine-bearing molecule in space after HF \citep[e.g.,][]{2005ApJ...628..260N,2006A&A...454L..37N}. Based on the fluorine chemical network, it has been suggested that 
CF$^{+}$ primarily forms through the reaction C$^{+} + $HF$\to$CF$^{+} +$H \citep{2005ApJ...628..260N,2019ApJ...872..203D}. Consequently, CF$^{+}$ is expected to be produced in PDRs where C$^{+}$ and HF are abundant \citep[e.g.,][]{2006A&A...454L..37N}. Therefore, CF$^{+}$ can serve as a proxy for C$^{+}$ and HF, which is supported by previous observations \citep[e.g.,][]{2012A&A...543L...1G}.



Since its first detection in the Orion Bar, an archetypal dense PDR \citep{2006A&A...454L..37N}, CF$^{+}$ has been widely detected in various Galactic regions including other dense PDRs \citep{2012A&A...548A..94G,2012A&A...543L...1G,2013A&A...550A..96N,2017A&A...599A..22N}, high-mass star formation regions \citep{2010ARep...54..295K,2010ARep...54.1084K,2015A&A...574L...4F,2015A&A...579A..12L}, and diffuse molecular clouds against bright background continuum sources \citep{2014A&A...564A..64L,2015A&A...579A..12L,2019A&A...622A..26G}. Recent imaging observations of CF$^{+}$ (2$\to$1) have revealed its extended distribution in Orion A \citep{2020A&A...636A..39B}. However, extragalactic studies of CF$^{+}$ have remained limited, primarily due to its weak intensities, with only one report of the CF$^{+}$ (2$\leftarrow$1) line in absorption along the line of sight toward the gravitationally lensed system PKS~1830$-$211 at a redshift $z=0.89$ \citep{2016A&A...589L...5M}. Hence, extragalactic CF$^{+}$ has remained largely unexplored. This raises the question of whether CF$^{+}$ can serve as a reliable probe for C$^{+}$ and fluorine abundances in external galaxies.

At a distance of 50 kpc \citep{Piet2019}, the Large Magellanic Cloud (LMC) stands out as one of the nearest metal-poor galaxies. Widespread bright [CII] emission from the LMC's ISM has been found \citep[e.g.,][]{1996ApJ...465..738I,2019A&A...621A..62O,2019MNRAS.490.3909O,2019A&A...632A.106L} with elevated C$^+$-to-CO ratios \citep{1995ApJ...454..293P}. As to fluorine, observations of the $2.3~\mu$m near infrared absorption band from HF have yielded fluorine abundances for a sample of red giants in the LMC \citep{2003AJ....126.1305C}, while fluorine in the system's ISM remains unconstrained.
These findings motivate searches for 
CF$^{+}$. 
Within the LMC, N113 emerges as a prominent site of active massive star formation, characterized by prolific molecular line emissions \citep{1997A&A...317..548C,1998A&A...332..493H,2009ApJ...690..580W,2014A&A...572A..56P,2016ApJ...818..161N,2018ApJ...853L..19S}, the presence of OH and H$_{2}$O masers \citep{1997MNRAS.291..395B,2010MNRAS.404..779E,2013MNRAS.432L..16I}, and numerous young stellar objects \citep[e.g.,][]{2009ApJS..184..172G,2010A&A...518L..73S,2012A&A...542A..66C}. Moreover, N113 ranks among the brightest [CII] emitters within the LMC \citep{2019MNRAS.490.3909O}, making it an exceptional candidate for the quest to detect CF$^{+}$. Therefore, we have undertaken dedicated observations of CF$^{+}$ towards N113, aiming to assess its role in the LMC's interstellar chemistry and as a tracer of C$^+$. 

We describe our observations in Sect.~\ref{Sec:obs}, and report our results in Sect.~\ref{Sec:res}. These findings are discussed in Sect.~\ref{Sec:dis}. We conclude with a summary in Sect.~\ref{Sec:sum}.

\section{Observations and data reduction}\label{Sec:obs}
On 2023 June 18--22 and August 22--26, targeted observations of N113 were conducted utilizing the Atacama Pathfinder EXperiment 12~m submillimeter telescope \citep[APEX;][]{2006A&A...454L..13G} under project code: M9514C\_111. For these observations, we employed the SEPIA180\footnote{The multi-module Swedish-ESO PI Instrument for APEX (SEPIA) was designed and built by the Group for Advanced Receiver Development (GARD), at Onsala Space Observatory (OSO) in collaboration with ESO; see\url{https://www.apex-telescope.org/ns/instruments/sepia/sepia180/}} and SEPIA345\footnote{\url{https://www.apex-telescope.org/ns/instruments/sepia/sepia345/}} receivers \citep{2018A&A...611A..98B,2018A&A...612A..23B,2022A&A...668A...2M}, equipped with an advanced generation Fast Fourier Transform Spectrometer \citep[FFTS;][]{2012A&A...542L...3K}, to observe the $J=2\to 1$ and $J=3\to 2$ transitions of CF$^{+}$ at $\sim$205.2~GHz and $\sim$307.7~GHz (more accurate values are presented in Table~\ref{Tab:lin}), respectively. The SEPIA180 and SEPIA345 frequency configurations provided instantaneous intermediate frequency (IF) bandwidths of 8 GHz and 16 GHz, respectively. Data processing was facilitated by FFTSs, which encompassed the aforementioned IF bandwidth using interconnected 4 GHz wide modules, each equipped with 65,536 channels. This configuration resulted in a channel width of 61 kHz, corresponding to $\sim$0.09~\kms\,and $\sim$0.06~\kms, respectively at 205.2 and 307.7 GHz, the rest frequencies of the CF$^+$ transitions; see Table \ref{Tab:lin} for accurate values. 

We conducted the observations using the APEX Control System (APECS) in the position-switching mode \citep{2006A&A...454L..25M}. Right ascension and declination of our target position are 
$(\alpha_{\rm J2000}, \delta_{\rm J2000}$) =(05$^{\rm h}$13$^{\rm m}$17\rlap{.}$^{\rm s}$40, $-$69\degree22\arcmin22\rlap{.}\arcsec0), which is the same as those used in \citet{2023A&A...679L...6G}. Calibration was performed at intervals of roughly every five minutes. System temperatures were 101--159~K and 103--195~K on a $T_{\rm A}^{*}$ scale at 205~GHz and 307~GHz, respectively. We employed a main beam efficiency of 84\%\,for both the $J$=2$\to$1 and $J$=3$\to$2 transitions according to pointing measurements toward Mars\footnote{\url{http://www.apex-telescope.org/telescope/efficiency/}}, to establish the main beam brightness temperature scale. The half-power beam widths (HPBWs), $\theta_{\rm b}$, are 29\arcsec\,and 19\arcsec\,at 205.2~GHz and 307.7~GHz, respectively. 

Spectral data reduction was undertaken using the Grenoble Image and Line Data Analysis Software  (GILDAS\footnote{\url{https://www.iram.fr/IRAMFR/GILDAS/}}; \citealt{2005sf2a.conf..721P}). To enhance the signal-to-noise ratios, we smoothed the observed CF$^{+}$ spectra to achieve a channel spacing of 1~\kms\,for our subsequent analysis.

\begin{table*}[!hbt]
\caption{Observed properties of the CF$^{+}$ transitions.}\label{Tab:lin}
\normalsize
\centering
\begin{tabular}{cccccccc}
\hline \hline
line             & Frequency      & $A_{ij}$  & $\theta_{\rm b}$  & $\varv_{\rm LSR}$ & $T_{\rm p}$  & $\Delta \varv$ & $\int T_{\rm mb} {\rm d}\varv$  \\ 
                 & (MHz)          &  (s$^{-1}$) &  (\arcsec)   & (\kms)    & (mK)   & (\kms)    & (mK~\kms) \\ 
(1)              & (2)            &  (3)        & (4)          & (5)       & (6)    & (7)       &  (8)      \\
\hline
CF$^{+}$ $J=2\to 1$ & 205170.52(2) & 4.6$\times 10^{-5}$ & 29 & 235.0$\pm$0.4 & 8.1$\pm$2.0 & 4.7$\pm$0.8 & 40.2$\pm$6.6   \\
CF$^{+}$ $J=3\to 2$ & 307744.35(2) & 1.7$\times 10^{-4}$ & 19 & 234.9$\pm$0.5 & 6.2$\pm$1.6   & 7.1$\pm$1.1 & 46.6$\pm$6.2   \\
\hline
\end{tabular}
\tablefoot{(1) Transition. (2) Rest frequency measured by \citet{Cazzoli2010}, taken from the Cologne Database for Molecular Spectroscopy \citep[CDMS\footnote{https://zeus.ph1.uni-koeln.de/cdms},][]{2016JMoSp.327...95E}. Uncertainties in the last decimal digit, suggested by the CDMS catalog, are given in parentheses. (3) Einstein A coefficient. (4) Half-power beam width. (5) Velocity centroid. (6) Peak main beam brightness temperature. (7) FWHM line width. (8) Integrated intensity. }
\normalsize
\end{table*}

\section{Results}\label{Sec:res}
\begin{figure}[!htbp]
\centering
\includegraphics[width = 0.49 \textwidth]{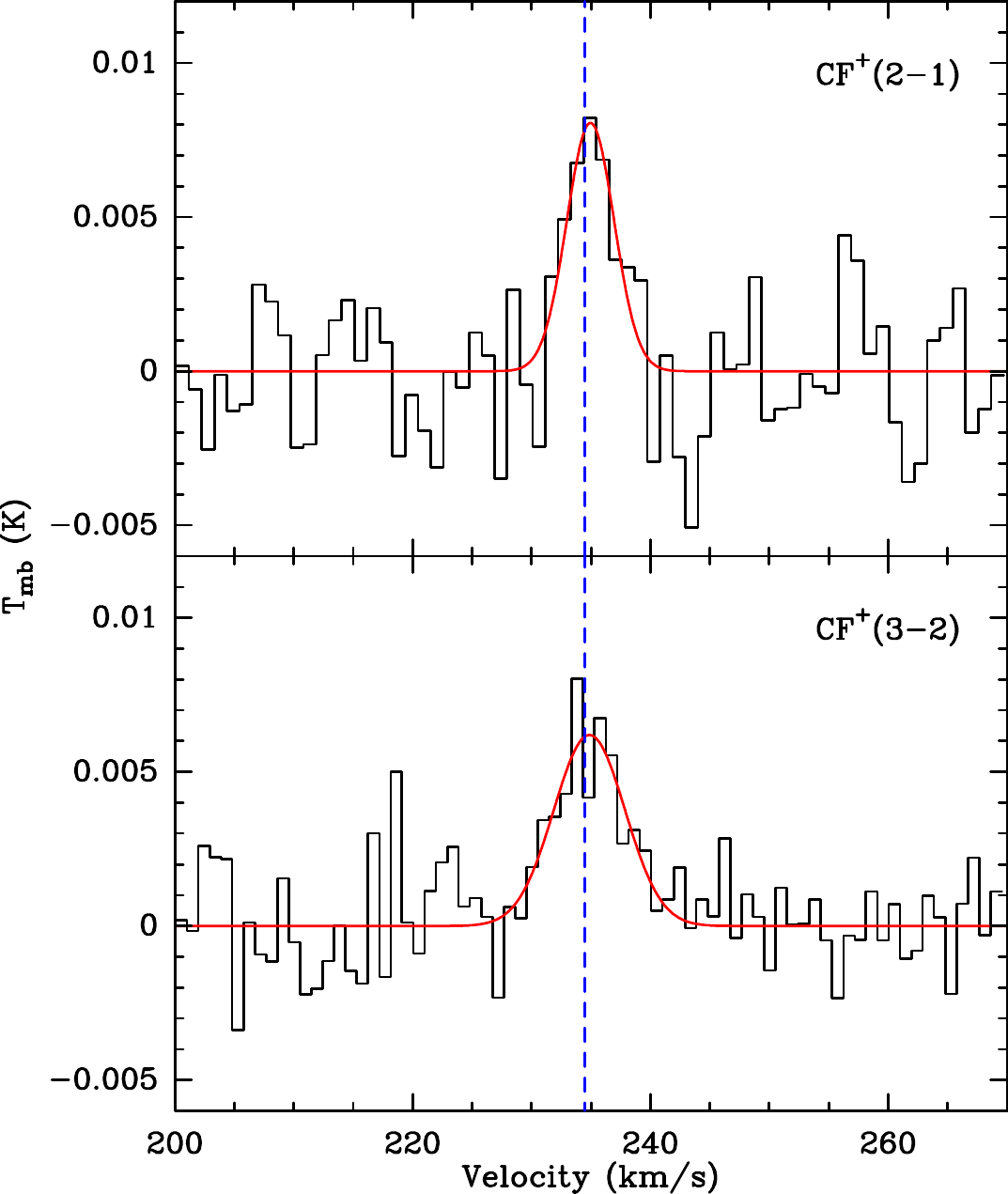}
\caption{{Observed transitions of CF$^{+}$ toward N113. The corresponding transition is indicated in the top right corner of each panel. The blue dashed line indicates the systemic velocity of N113. Both spectra have been smoothed to have a channel spacing of $\sim$1~\kms.}\label{Fig:spec}}
\end{figure}

In Fig.~\ref{Fig:spec}, we present the CF$^{+}$ spectra observed toward N113. Thanks to the high sensitivity of our observations, both spectra exhibit signal-to-noise ratios of $>$3 channel by channel and $>6$ for their integrated fluxes, affirming our successful detection of CF$^{+}$ toward N113. This is the first discovery of this molecular ion in the LMC, and the second extragalactic detection after that in the $z = 0.89$ foreground galaxy of the gravitational lens-magnified quasar PKS~1830$–$211 \citep{2016A&A...589L...5M}. CF$^{+}$ was detected in absorption toward PKS~1830$–$211, making our observations the first detection of CF$^{+}$ in emission in an extragalactic object. 

The CF$^{+}$ transitions exhibit splitting by hyperfine structure (HFS) \citep[see][who list the frequency separations of the HFS components]{2012A&A...548A..94G}. However, the corresponding velocity separations ($<$1~\kms) 
are too small to be resolved in our spectra. 
We find that a single Gaussian component provides adequate fits to our spectra, which delivers the  parameters listed in Table~\ref{Tab:lin}. The peak intensities are a factor of $>$10 lower than those observed in the Orion Bar \citep{2006A&A...454L..37N} and the Horsehead PDR \citep{2012A&A...543L...1G}. The measured LSR velocity centroids ($\approx$235~\kms) agree well with previous measurements of N113's systemic velocity 
\citep[e.g.,][]{2009ApJ...690..580W,2012ApJ...751...42S,2016ApJ...818..161N}. Taking the errors into account, we find consistent FWHM line widths for the two CF$^{+}$ transitions at the 2$\sigma$ level. The FWHM line widths are also akin to those of other dense gas tracers observed toward N113 \citep[e.g.,][]{2009ApJ...690..580W,2023A&A...679L...6G}, further corroborating our discovery of CF$^{+}$ in this source. Particularly, previous measurements indicate a line width of 4.78$\pm$0.23~\kms\,for the $^{13}$CO $J = 3 \to 2$ line measured with an angular resolution of 16\arcsec\,\citep{2009ApJ...690..580W}, which is comparable to the line widths observed for the CF$^{+}$ transitions. This similarity suggests that both the $^{13}$CO $J = 3 \to 2$ and CF$^{+}$ lines delineate gas undergoing identical turbulent motions. The observed FWHM line widths of CF$^{+}$ appear to be broader than those typically observed toward Galactic sources \citep[$\lesssim$3~\kms;][]{2006A&A...454L..37N,2012A&A...548A..94G,2012A&A...543L...1G,2013A&A...550A..96N,2017A&A...599A..22N} but narrower than that observed toward PKS~1830$-$211 \citep[$\sim$15~\kms;][]{2016A&A...589L...5M}. 
This difference likely arises from the different linear scales probed by these various studies.

In light of the direct link between CF$^{+}$ and C$^{+}$ \citep[e.g.,][]{2005ApJ...628..260N,2006A&A...454L..37N,2012A&A...543L...1G}, we reanalyzed the \textit{Herschel} [CII] 158~$\mu$m emission data of N113 which was previously reported by \citet{2019MNRAS.490.3909O}. To allow for a meaningful comparison, we convolved the Herschel [CII] data to match the angular resolutions of our CF$^{+}$ data. The resulting [CII] integrated intensities are 113.9$\pm$0.4~K~\kms\,and 98.4$\pm$0.3~K~\kms\,at the angular resolutions of 19\arcsec\,and 29\arcsec, respectively. These values exceed those of the CF$^{+}$ transitions by over three orders of magnitude in brightness temperature.



\begin{figure}[!htbp]
\centering
\includegraphics[width = 0.49 \textwidth]{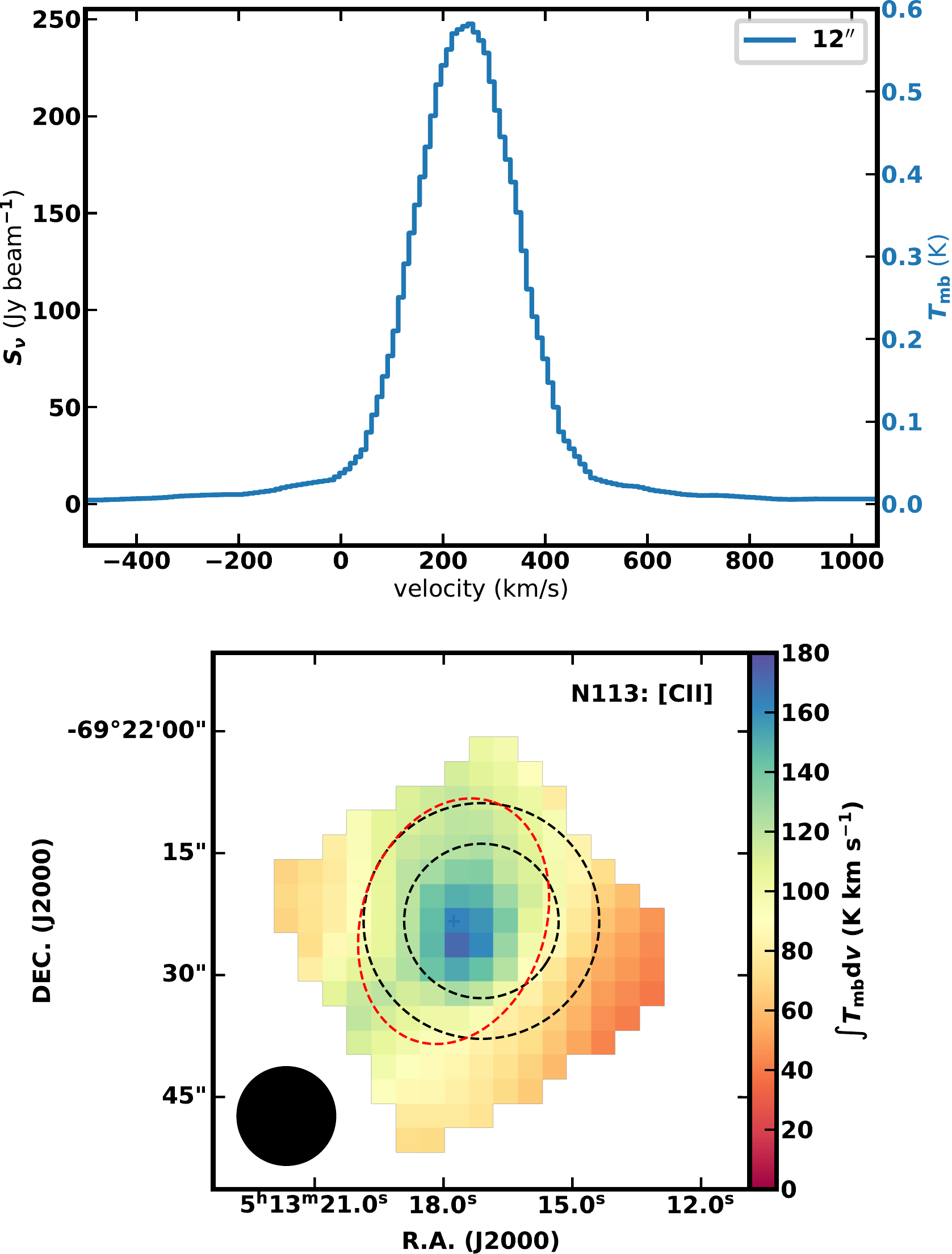}
\caption{{\textit{Top}: \textit{Herschel}/PACS [CII] spectra of the target position. We note that the line profiles are predominantly broadened by the spectral resolution of \textit{Herschel}/PACS, $\lambda/\Delta \lambda \sim 1500$ (i.e., $\sim$200~\kms). A flux uncertainty of 10\%\,is assumed for this measurement \citep[see Table A1 in][]{2019MNRAS.490.3909O}.
\textit{Bottom}: \textit{Herschel} continuum-subtracted [CII] integrated intensity map at its original angular resolution of 12\arcsec. The two black dashed circles represent the APEX HPBWs of the CF$^{+}$ (2$\to$1) and (3$\to$2) observations, while the Gaussian fitting to the [CII] distribution is indicated by the red dashed ellipse. The 158~$\mu$m \textit{Herschel} HPBW is shown in the lower left.}\label{Fig:CII}}
\end{figure}

Given the limitation of our pointed observations, the spatial distribution of CF$^{+}$ emission remains unconstrained by our measurements. However, as C$^{+}$ is the main precursor of CF$^{+}$, we anticipate that the distribution of CF$^{+}$ emission might follow that of C$^{+}$. The bottom panel of Fig.~\ref{Fig:CII} shows the distribution of the continuum-subtracted [CII] integrated intensity map of N113. 
Our observed position coincides with the bright [CII] emission in N113 \citep{2019MNRAS.490.3909O}, but is about 3\rlap{.}\arcsec5 offset with respect to the [CII] peak. The [CII] distribution shows a compact source superimposed on a more extended emission. We performed a Gaussian fitting to the integrated intensity map, yielding a FWHM size of 22\rlap{.}\arcsec6$\times$30\rlap{.}\arcsec9 with a position angle of 72\degree. This is equivalent to the deconvolved Gaussian FWHM size, $\theta_{\rm s}$, of 23\rlap{.}\arcsec5 (i.e., 6~pc) for the [CII] emission. 
Since the CF$^{+}$ transitions are assumed to trace high-density gas (see Sect.~\ref{sec.den}), we adopt $\theta_{\rm s}$=23\rlap{.}\arcsec5 as an upper limit for our subsequent analysis.

Assuming an excitation temperature of 10~K \citep[e.g.,][]{2006A&A...454L..37N,2012A&A...543L...1G} and a Gaussian FWHM source size of 23\rlap{.}\arcsec5, we estimate peak optical depths of 0.004 and 0.002 for CF$^{+}$ (2$\to$1) and (3$\to$2), respectively, based on their peak intensities ($<$10~mK). Even when considering lower excitation temperatures (i.e., 5~K) and smaller source sizes of 10\arcsec--20\arcsec, the optical depths remain below 0.1. Furthermore, our statistical equilibrium calculations also confirm that the optical depths are lower than 0.1 (see Sect.~\ref{sec.den}). These findings strongly support that both CF$^{+}$ transitions can be reliably taken to be optically thin. 

\begin{figure}[!htbp]
\centering
\includegraphics[width = 0.49 \textwidth]{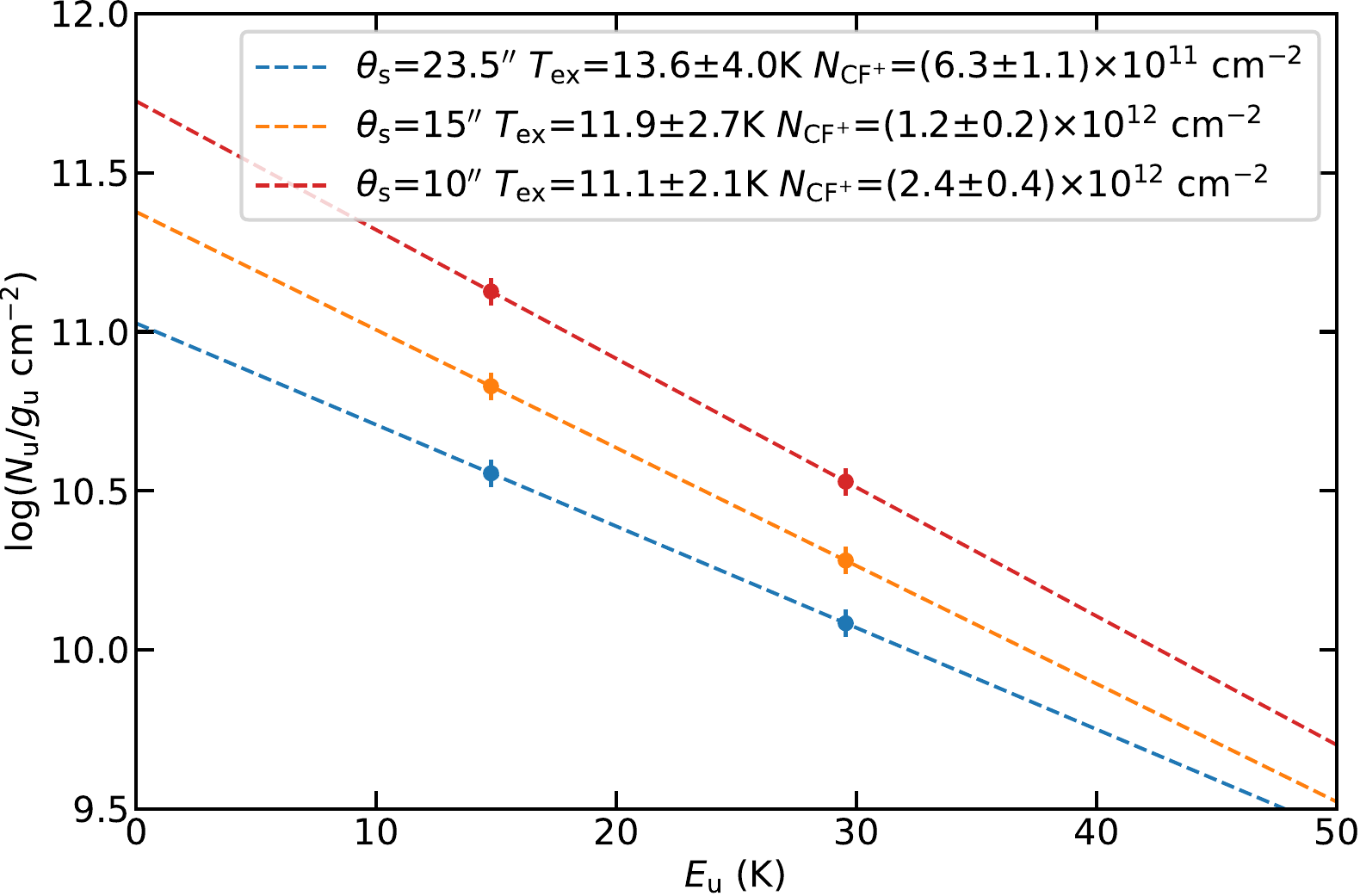}
\caption{{Rotational diagram of CF$^{+}$ in N113. The dashed lines represent the linear least-square fits to the observed data points at three assumed source sizes. The adopted source size, fitted excitation temperature, and CF$^{+}$ column density are shown in the upper right.}\label{Fig:rd}}
\end{figure}

To estimate the excitation temperature and column density of CF$^{+}$, we employed the rotational diagram, which assumes conditions of Local Thermodynamic Equilibrium (LTE) \citep[e.g.,][]{1999ApJ...517..209G}, where the intensities were corrected by dividing by the beam dilution factor, $\frac{\theta_{\rm s}^{2}}{\theta_{\rm s}^{2}+\theta_{\rm b}^{2}}$. Because we lack precision source size constraints for the CF$^{+}$ emitting region in N113, we made calculations assuming three different source sizes (i.e., $\theta_{\rm s}$=23\rlap{.}\arcsec5, 15\arcsec, and 10\arcsec). The results are illustrated in Fig.~\ref{Fig:rd}. Utilizing linear fits to the rotational diagram, we obtain rotation temperatures of 13.6$\pm$4.0~K, 11.9$\pm$2.7~K, 11.1$\pm$2.1~K, and CF$^{+}$ column densities of (6.3$\pm$1.1)$\times 10^{11}$~cm$^{-2}$, (1.2$\pm$0.2)$\times 10^{12}$~cm$^{-2}$, and (2.4$\pm$0.4)$\times 10^{12}$~cm$^{-2}$ for source sizes of 23\rlap{.}\arcsec5, 15\arcsec, and 10\arcsec, respectively. The rotation temperatures, which, under LTE conditions, are identical to the excitation temperatures, remain relatively consistent in the range of 10--15~K and do not vary significantly with the assumed source size. These values are reasonably in line with previous studies of CF$^+$ in the Milky Way \citep[$\sim$10~K;][]{2006A&A...454L..37N,2012A&A...543L...1G}, but are slightly lower than the excitation temperature of 22$\pm$4~K determined by \citet{2017A&A...599A..22N} for the Orion Bar. The CF$^{+}$ column densities determined toward N113 are not significantly different from the values of $(1-2)\times 10^{12}$~cm$^{-2}$ estimated toward the Orion Bar and the Horsehead PDR \citep{2006A&A...454L..37N,2012A&A...543L...1G,2017A&A...599A..22N}. 
This implies that CF$^{+}$ column densities exhibit a level of consistency within an order of magnitude across various PDRs. 

\section{Discussion}\label{Sec:dis}
\subsection{Density constraints}\label{sec.den}
Previous investigations have demonstrated the utility of multiple CF$^{+}$ transitions in deriving gas densities within the Milky Way \citep{2021A&A...645A...8D}. Our successful detection of two CF$^{+}$ transitions allows us to extend such analyses to an extragalactic environment for the first time. To constrain the gas density of the CF$^{+}$ emitting zone in N113, we used the RADEX\footnote{\url{https://home.strw.leidenuniv.nl/~moldata/radex.html}} code for statistical equilibrium calculations \citep{2007A&A...468..627V}. Our analysis made use of the molecular data for CF$^{+}$ from the Leiden Atomic and Molecular Database \citep[LAMDA\footnote{\url{https://home.strw.leidenuniv.nl/~moldata/}};][]{2005A&A...432..369S}. Specifically, energy levels, rest frequencies, and Einstein A coefficients were extracted from the CDMS catalog \citep{2016JMoSp.327...95E}. Additionally, we adopted the latest rate coefficients for collisions of CF$^{+}$ with H$_{2}$ from \citet{2021A&A...645A...8D}. A total of 22 energy levels and 21 rotational transitions were taken into account in our calculations. Similar to the approach of \citet{2021A&A...645A...8D}, our work exclusively involves collisions of CF$^{+}$ with para-H$_{2}$. This is because the computed rate coefficients show no dependence on the nuclear spin (see \citealt{2021A&A...645A...8D} for information). We also note that electron excitation might play a role. However, the CF$^{+}$-e$^{-1}$ collisional rates are not available. Adopting the Born approximation formula, as in \citet{1989PhRvA..40..633N}, we estimate that the CF$^{+}$-e$^{-1}$ collisional rate coefficients are of the order of $10^{-7}$~cm$^{-3}$~s$^{-1}$. These values significantly exceed the corresponding rate coefficients for collisions with para-H$_{2}$, which are of the order of $10^{-10}$~cm$^{-3}$~s$^{-1}$. Assuming electron abundances comparable to the average carbon abundance relative to H in the LMC \citep[$8\times 10^{-5}$;][]{1982ApJ...252..461D}, the electron number density is lower than that of para-H$_{2}$ by about four orders of magnitude. Furthermore, the electron abundances are expected to be even lower in dense regions like N113 \citep[e.g.,][]{2002ApJ...565..344C}. Hence, we suggest that electron collisions with CF$^{+}$ are negligible in our calculations.  


In the modeling process, we constructed regular grids with kinetic temperatures, $T_{\rm k}$, ranging from 5~K to 150~K with a step of 1~K, and H$_{2}$ number densities, $n$(H$_{2}$), ranging from $10^{3}$~cm$^{-3}$ to $10^{7}$~cm$^{-3}$ with a logarithmic step of log[$n$(H$_{2}$)~/cm$^{-3}$] =0.1. The column densities and source sizes are based on our observations and the LTE analysis (see Sect.~\ref{Sec:res}), while the line widths are fixed to be 4.7~\kms (see Table~\ref{Tab:lin}). The physical parameters were optimized by minimizing the $\chi^{2}$ value, which is defined as:
\begin{equation}\label{f.chi}
    \chi^{2} = \Sigma_{i} \frac{(I_{\rm obs,i}-I_{\rm mod, i})}{\sigma_{i}} \;,
\end{equation}
where $I_{\rm obs,i}$ and $I_{\rm mod, i}$ represent the observed and modeled integrated intensities, respectively, and $\sigma_{i}$ denotes the error in $I_{\rm obs,i}$. In Eq.~(\ref{f.chi}), $I_{\rm obs,i}$ has been corrected for the beam dilution effect. 

\begin{figure*}[!htbp]
\centering
\includegraphics[width = 0.95 \textwidth]{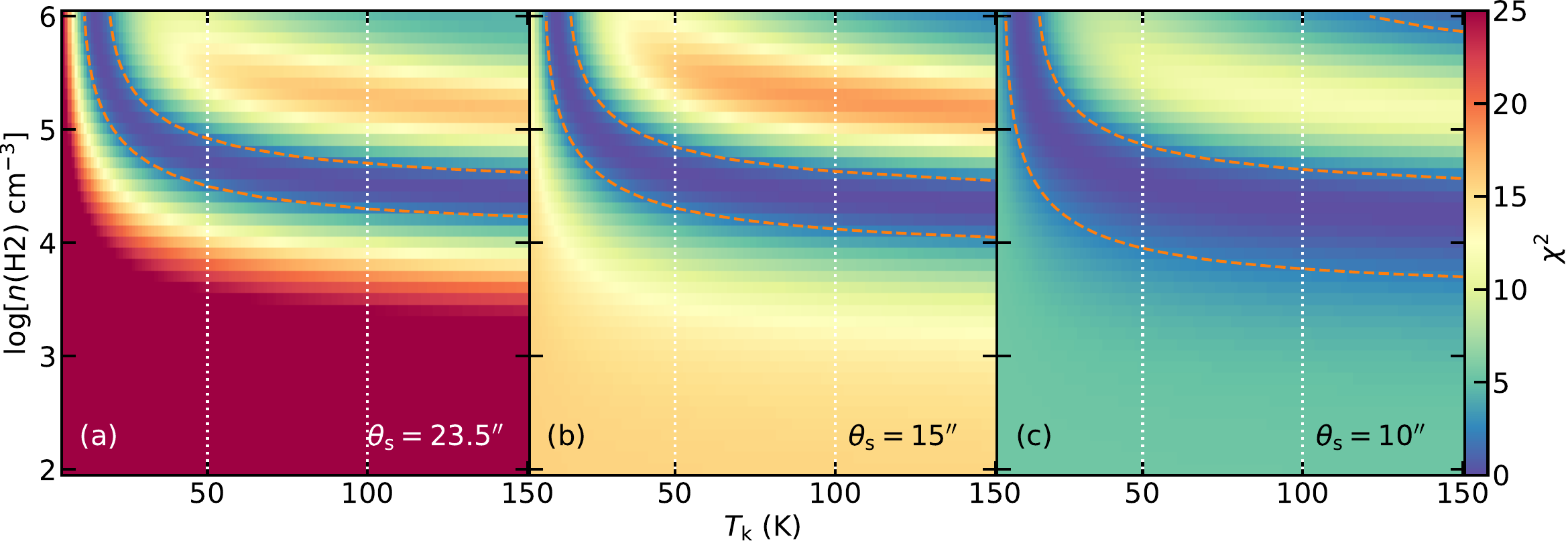}
\caption{{Distribution of $\chi^{2}$ as a function of the kinetic temperature and H$_{2}$ number density. The adopted source size is shown in the lower right corner of each panel. The orange dashed lines mark the range of $\chi^{2}<$2.3 that corresponds to the 1$\sigma$ significance level. The white vertical dashed lines enclose the expected kinetic temperature range.}\label{Fig:radex}}
\end{figure*}

Figure~\ref{Fig:radex} presents the distribution of ${\chi^{2}}$ as a function of the kinetic temperature and H$_{2}$ number density from our modeling results. As is clear from this figure, the kinetic temperature is not well constrained by our modeling. However, the H$_{2}$ number density does not vary much for kinetic temperatures above 30~K. Previous H$_{2}$CO measurements of N113 suggest kinetic temperatures of $\sim$50--100~K  \citep{2017A&A...600A..16T,2021A&A...655A..12T}, providing a basis for inferring the H$_{2}$ number density in the CF$^{+}$ emitting region of N113. Using the criterion of $\chi^{2}<$2.3, corresponding to the 1$\sigma$ confidence level, and assuming a prior on kinetic temperatures of $T_{\rm k}\sim$50--100~K, we infer H$_{2}$ number densities in the range of $10^{4.3-4.9}$~cm$^{-3}$, $10^{4.1-4.8}$~cm$^{-3}$, and $10^{3.7-4.8}$~cm$^{-3}$ for source sizes of 
23\rlap{.}\arcsec5, 15\arcsec, and 10\arcsec, respectively. While the derived number densities show a slight decrease with diminishing source size, overall, they do not vary significantly. Hence, we conclude that CF$^{+}$ should emerge from molecular gas with an H$_{2}$ number density of $(0.5-7.9)\times 10^{4}$~cm$^{-3}$.

When compared with previous density estimates for the Horsehead PDR and the Orion Bar from CF$^{+}$ transitions \citep{2021A&A...645A...8D},
the H$_{2}$ number density of N113 appears to be comparable to that of the Horsehead PDR ($\sim 3.5\times 10^{4}$~cm$^{-3}$) , but slightly lower than that of the Orion Bar ($\sim 1.8\times 10^{5}$~cm$^{-3}$). Toward N113, the derived H$_{2}$ number densities appear to be lower than those ($>$1$\times 10^{5}$~cm$^{-3}$) derived from dense gas tracers like CS, HCN, HCO$^{+}$, H$_{2}$CO, and HC$_{3}$N at similar angular resolutions \citep[see Table~5 in][]{2009ApJ...690..580W}, indicating that our CF$^{+}$ transitions probe lower density molecular gas.  

\subsection{Fluorine abundances}\label{sec.fluorine}
Fluorine abundances have been studied utilizing both neutral fluorine \citep{2005ApJ...619..884F,2007ApJ...655..285S} and ionized fluorine \citep{2005A&A...433..641W,2005ApJ...631L..61Z}. However, the main method employed for measurements remains centered on the HF molecule, given its significance as the primary fluorine carrier \citep[e.g.,][]{MATHEW2023110149}.
Consequently, observations of ro-vibrational HF absorption serve as a common approach for investigating fluorine abundances in cool stars within the Milky Way and the LMC \citep[e.g.,][]{1992A&A...261..164J,2003AJ....126.1305C,2005ApJ...626..425C}. Employing this methodology, the fluorine abundances determined for the Orion K–M dwarfs are comparable to the Solar value \citep{2005ApJ...626..425C}. In molecular clouds, assessments of fluorine abundances predominantly rely on observations of pure rotational transitions of HF in the vibrational ground state. However, the commonly used (1$\to$0) rotational transition of HF, with a rest frequency of $\sim$1232~GHz, has a much higher critical density\footnote{HF (1$\to$0) has a critical density of $\sim$5$\times 10^{9}$~cm$^{-3}$ \citep{2012A&A...537L..10V}, while low $J$ CF$^{+}$ lines have critical densities of $<2\times 10^{6}$~cm$^{-3}$ \citep[see Table 1 in][]{2021A&A...645A...8D}.} than the CF$^+$ lines studied here and is observed mostly in absorption that can become optically thick \citep{2010A&A...518L.108N}. This introduces uncertainties in estimations of fluorine abundances based on HF in many cases. Moreover, as mentioned above, observations of these far-infrared rotational lines are currently not possible due to the absence of facilities operating at these far-infrared wavelengths. In contrast, CF$^{+}$ transitions can be readily observed by (sub)millimeter telescopes and typically exhibit low optical depths. Therefore, observations of CF$^{+}$ transitions can offer an alternative method for fluorine abundance determinations.

According to previous studies \citep{2005ApJ...628..260N,2006A&A...454L..37N,2012A&A...543L...1G}, CF$^{+}$ is formed through the reactions:
\begin{equation}
    {\rm F + H_{2}} \to {\rm HF + H}
\end{equation}
\begin{equation}\label{f.form}
    {\rm HF + C^{+}} \xrightarrow{k_{1}} {\rm CF^{+} + H}
\end{equation}
and mainly destroyed by dissociative recombination with electrons:
\begin{equation}\label{f.des}
 {\rm CF^{+} + e^{-}} \xrightarrow{k_{2}} {\rm C+ F}\;.
\end{equation}
This simple chemical network provides a straightforward link between fluorine abundances, $[{\rm F}]$, CF$^{+}$ column densities, $N({\rm CF^{+}})$, and total hydrogen column densities, $N_{\rm tot}({\rm H})$, through the expression \citep{2012A&A...543L...1G}:
\begin{equation}\label{f.relation}
N({\rm CF^{+}}) = \frac{k_1}{k_2} [{\rm F}] N_{\rm tot}({\rm H})\;,
\end{equation}
where $k_1$ and $k_2$ are the rate coefficients of reactions~(\ref{f.form}) and (\ref{f.des}), respectively. 

Equation~(\ref{f.relation}) highlights the dependence of [F] on the ratio $\frac{k_{1}}{k_{2}}$. 
The most recent calculations for $k_{1}$ were performed by \citet{2019ApJ...872..203D}, reporting $k_{1}=4.33\times 10^{-9}(T_{\rm k}/300)^{-0.425}{\rm exp}(-18.29/T_{\rm k})$~cm$^{3}$~s$^{-1}$ and $1.50\times 10^{-9}(T_{\rm k}/300)^{-0.442}{\rm exp}(-8.84/T_{\rm k})$~cm$^{3}$~s$^{-1}$ for the $j$=1/2 and $j$=3/2 spin-orbit states, respectively. Assuming that the majority of C$^{+}$ exists in the lower energy state, we adopt $k_1$ of the $j$=1/2 spin-orbit state for our analysis. Based on measurements from a storage ring experiment, \citet{Novotny_2005} found $k_2 =5.30\times 10^{-8}(T_{\rm k}/300)^{-0.8}$~cm$^{3}$~s$^{-1}$. Based on previous H$_{2}$CO measurements, the kinetic temperatures of N113 are expected to fall within the range of 50--100~K \citep{2017A&A...600A..16T,2021A&A...655A..12T}, which leads to $\frac{k_{1}}{k_{2}}$ ranging from 0.029 to 0.045.  

Based on the results in Appendix~\ref{app.a}, we adopted $N_{\rm tot}({\rm H})=(5.8\pm 0.6)\times 10^{22}$~cm$^{-2}$. Since our observations suggest $N({\rm CF^{+}})=(0.6-2.5)\times 10^{12}$~cm$^{-2}$ (see Sect.~\ref{Sec:res}), we obtain fluorine abundances within the range of [F]=$(0.2-1.7)\times 10^{-9}$ using Eq.~(\ref{f.relation}). Because the adopted total hydrogen column densities are derived at an angular resolution of $\sim$40\arcsec, these values are likely lower than total hydrogen column densities determined for higher angular resolution data. Thus, the derived fluorine abundances are considered as upper limits (i.e., [F]$\lesssim 1.7\times 10^{-9}$). Given that fluorine abundances in molecular clouds within the LMC have not been previously measured, our results represent the first constraint on fluorine abundances in such environments in the LMC. 

The measured fluorine abundance in N113 is lower than the value in the Solar system \citep[$3\times 10^{-8}$;][]{1989GeCoA..53..197A}, but comparable to the abundance ($\sim2\times 10^{-9}$) observed in a gravitationally lens-magnified dusty star-forming galaxy at redshift $z=4.4$ \citep{2021NatAs...5.1240F}. Because fluorine is mainly produced by the helium burning process of asymptotic giant branch (AGB) stars \citep[e.g.,][]{1992A&A...261..164J,2005A&A...433..641W}, the low fluorine abundance in the lensed galaxy was attributed to insufficient enrichment by AGB stars in the early Universe. However, this explanation does not apply to the LMC, given its considerably older age \citep[e.g.,][]{2009AJ....138.1243H}, which is sufficient for enrichment by AGB stars. On the other hand, the measured fluorine abundance in N113 is about an order of magnitude lower than those ([F]$\sim 10^{-8}$) measured toward red giants in the LMC \citep{2003AJ....126.1305C}, indicating a substantial deficiency in N113. Based on previous measurements of CF$^{+}$ toward the Orion Bar \citep{2017A&A...599A..22N}, we determined a $N({\rm CF}^{+})/N({\rm H})$ ratio of about $3.1\times 10^{-11}$, corresponding to [F]$\lesssim 1.1\times 10^{-9}$. Thus, the fluorine abundance in the Orion Bar is also about an order of magnitude lower than those ($\sim 3.8\times 10^{-8}$) measured toward stars in the Orion nebula cluster \citep{2005ApJ...626..425C}, confirming the trend observed in the LMC. 

Previous observations have shown that HF abundances sharply decline in dense environments \citep{2012ApJ...756..136E}. Similarly, HF abundances were estimated to be as low as a few $10^{-10}$ in massive star-formation regions like Sgr B2 and NGC6334I \citep{1997ApJ...488L.141N,2012ApJ...756..136E,2016A&A...593A..37V}, attributed to the freeze-out of HF molecules onto dust grains. Following Eqs.~(3) and (4) in \citet{2016A&A...593A..37V}, we find that the HF adsorption rate surpasses its desorption rate at a kinetic temperature of 100~K and an H number density of $\sim 2\times 10^{4}$~cm$^{-3}$, which indicates that the freeze-out of HF molecules can effectively take place in N113. As HF serves as the main precursor of CF$^{+}$, the freeze-out of HF molecules provides a natural explanation for the observed deficiency of CF$^{+}$ ions in N113. Therefore, our findings suggest that fluorine abundances measured in dense molecular clouds only represent lower limits for fluorine abundances in galaxies due to the observed CF$^{+}$ deficiency.

Our simplified network overlooks the potential impact of photodissociation on the destruction of HF and CF$^{+}$, which could in principle lead to an overestimate of HF and CF$^{+}$ abundances. However, both HF and CF$^{+}$ exhibit remarkable stability, characterized by dissociation energies of 5.87~eV and 7.71~eV \citep{2006A&A...454L..37N}, respectively. Based on the KInetic Database for Astrochemistry (KIDA\footnote{\url{https://kida.astrochem-tools.org/}}), the photodissociation rate of HF is $1.38\times 10^{-10} {\rm exp}(-2.99A_{\rm V})$~s$^{-1}$, where $A_{\rm V}$ is the visual extinction. This indicates an exceedingly slow photodissociation rate of $\lesssim10^{-10}$. Based on the assumed photodissociation rate of $10^{-9} {\rm exp}(-2.5A_{\rm V})$ s$^{-1}$ for CF$^{+}$ \citep{2012A&A...543L...1G}, our calculations indicate that the photodissociation rate of CF$^{+}$ should be considerably lower than $10^{-9}$~s$^{-1}$. This finding is further supported by \citet{2015A&A...579A..12L}, who proposed a value of $2\times 10^{-10}$~s$^{-1}$. Given these low rates, we posit that the contribution of the photodissociation effects can be neglected. On the other hand, we note that the photodissociation rates are still uncertain. As discussed by \citet{2015A&A...579A..12L}, photodissociation might have the potential to impact CF$^{+}$ abundances if its rates exceed $\gtrsim 10^{-9}$~s$^{-1}$. Furthermore, the line-of-sight structures of PDRs can exhibit great complexity in reality, potentially permitting the penetration of UV radiation. Additionally, since CF$^{+}$ emissions likely trace only a fraction of H$_{2}$ gas, its abundance derived from the assumed hydrogen column density is underestimated. The photodissociation effects and the overestimated hydrogen column density may lead to lower values when compared with the fluorine abundance derived from our simplified network.


\subsection{C$^{+}$/CF$^{+}$ ratios}
Previous studies indicate that CF$^{+}$ could act as a proxy for C$^{+}$ \citep[e.g.,][]{2012A&A...543L...1G}. Establishing a known C$^{+}$/CF$^{+}$ ratio would enable the inference of C$^{+}$ column densities from CF$^{+}$ measurements. While the CF$^{+}$ column density of N113 has been investigated in Sect.~\ref{Sec:res}, the C$^{+}$ column density requires constraints derived from the \textit{Herschel}/PACS [CII] spectra (see Fig.~\ref{Fig:CII} for instance). 

Given that [CII] is expected to be optically thick in the LMC \citep{2019A&A...631L..12O}, adopting the optically thin approximation for estimating C$^{+}$ column densities may not be valid. Instead, we employed RADEX models to quantify the relation between the C$^{+}$ integrated intensity and its column density. The spectroscopic data for C$^{+}$ were adopted from the LAMDA database, encompassing 8 energy levels, 14 transitions of C$^{+}$, and incorporating the most recent collisional rate coefficients 
\citep{2005ApJ...620..537B,2002MNRAS.337.1027W,2013JChPh.138t4314L}. In our RADEX models, we considered two collisional partners, para-H$_{2}$ and H. Based on the estimate in Appendix~\ref{app.a}, we assumed $n({\rm H})/n({\rm H}_{2})=0.1$. The line width was held constant at 4.7~\kms, matching that of the CF$^{+}$ (2$\to$1) line. Similar to Sect.~\ref{sec.den}, we also constructed regular grids with kinetic temperatures ranging from 5~K to 150~K with a step of 1~K, H$_{2}$ number densities ranging from $10^{3}$~cm$^{-3}$ to $10^{7}$~cm$^{-3}$ with a logarithmic step of log[$n$(H$_{2}$)~cm$^{-3}$] =0.1, and C$^{+}$ column densities ranging from $10^{16}$~cm$^{-2}$ to $10^{19}$~cm$^{-2}$ with a logarithmic step of log[$N$(C$^{+}$)~/cm$^{-2}$] =0.1.  

\begin{figure}[!htbp]
\centering
\includegraphics[width = 0.49 \textwidth]{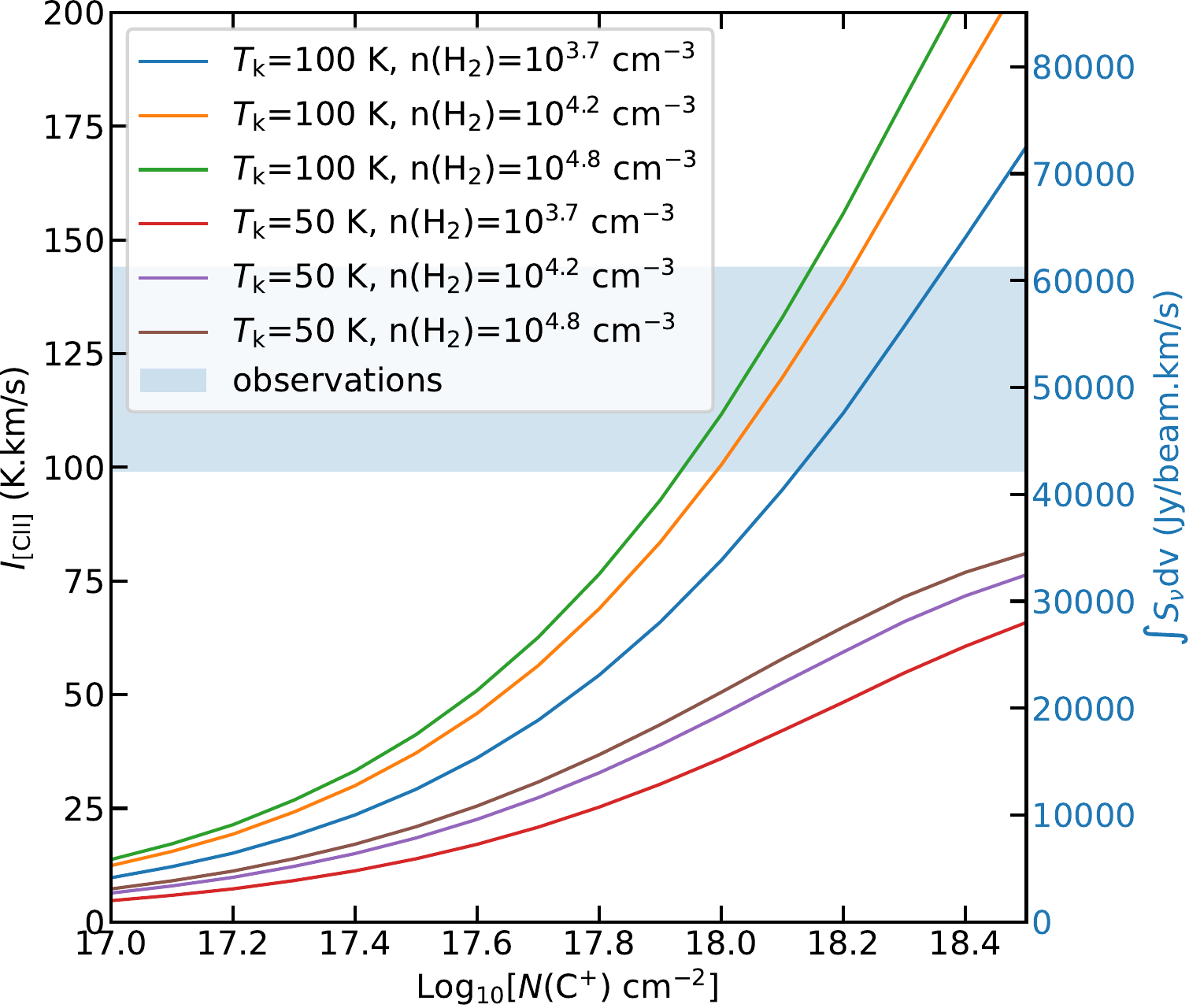}
\caption{{Relation between the C$^{+}$ column density and [CII] integrated intensity predicted by the RADEX models.}\label{Fig:model_c+}}
\end{figure}

Figure~\ref{Fig:model_c+} presents the relation between the C$^{+}$ column density and [CII] integrated intensity, calculated from the RADEX models under different physical conditions. Based on previous H$_{2}$CO observations of N113 \citep{2017A&A...600A..16T,2021A&A...655A..12T}, we investigate models with kinetic temperatures of $\sim$50--100~K. Since [CII] emission arises not only from molecular gas but also from ionized gas, the kinetic temperature of the [CII]-emitting region is anticipated to be higher than that of molecular gas traced by H$_{2}$CO. Furthermore, the low kinetic temperature of 50~K fails to reproduce the observed [CII] integrated intensities in Fig.~\ref{Fig:model_c+}. 
Hence, we adopt a kinetic temperature of 100~K to estimate the C$^{+}$ column density. A Gaussian fit to the spectra in Fig.~\ref{Fig:CII} gives [CII] integrated intensities within the range of 99.0--144.1~K~\kms\,for different beams of 12\arcsec--29\arcsec. Assuming the absence of beam dilution effects and comparing the observed values with the RADEX models (see Fig.~\ref{Fig:model_c+}), we derive a C$^{+}$ column density of 10$^{17.9-18.3}$~cm$^{-2}$ within the H$_{2}$ number density range determined in Sect.~\ref{sec.den}. The values are comparable to those found in the N159 star-forming region \citep[Table~2]{2015A&A...580A..54O}. If we assume that the emission from CF$^{+}$ is as extended as that from C$^{+}$ (i.e., $\theta_{\rm s}$=23\rlap{.}\arcsec5), the CF$^{+}$/C$^{+}$ column density ratio is about $(3.0-7.6)\times 10^{-7}$, which is comparable to the estimated value ($\sim5\times 10^{-7}$) in the Orion Bar \citep{2006A&A...454L..37N}. 

Based on Reactions~(\ref{f.form}) and (\ref{f.des}), \citet{2006A&A...454L..37N} suggested a theoretical C$^{+}$/CF$^{+}$ ratio in equilibrium,  
\begin{equation}\label{f.c+cf+}
    \frac{n({\rm CF}^{+})}{n({\rm C}^{+})} = \frac{k_{1}}{k_{2}}\frac{n({\rm HF})}{n_{\rm e}}\;,
\end{equation}
where $n_{\rm e}$ denotes the electron density. Assuming $n_{\rm e}\sim n({\rm C^{+}}) \sim n_{\rm C}$ and $n({\rm HF})\sim n_{\rm F}$, we can substitute Eq.~(\ref{f.c+cf+}) with 
\begin{equation}\label{f.c+cf+2}
    \frac{n({\rm CF}^{+})}{n({\rm C}^{+})} = \frac{k_{1}}{k_{2}}\frac{n_{\rm F}}{n({\rm H})}\frac{n({\rm H})}{n_{\rm C}} \;,
\end{equation}
where $\frac{n_{\rm F}}{n({\rm H})}$ represents the fluorine abundance, and $\frac{n_{\rm C}}{n({\rm H})}$ is the carbon abundance. Adopting $\frac{n_{\rm F}}{n({\rm H})}\sim 10^{-8}$ \citep{2003AJ....126.1305C} and $\frac{n_{\rm C}}{n({\rm H})}\sim 8\times 10^{-5}$ \citep{1982ApJ...252..461D} based on previous abundance measurements in the LMC, the theoretical $\frac{n({\rm CF}^{+})}{n({\rm C}^{+})}$ ratio is anticipated to be $(3.6-5.6)\times 10^{-6}$ within the temperature range of 50--100 K in the LMC, which is higher than our derived value of $(3.0-7.6)\times 10^{-7}$ in N113. As illustrated in Fig.~\ref{Fig:model_c+}, a decrease in kinetic temperature results in an elevated C$^{+}$ column density, thereby yielding even lower CF$^{+}$/C$^{+}$ column density ratios. Therefore, our conclusion that the derived CF$^{+}$/C$^{+}$ ratio is lower than the theoretical prediction remains robust even if we assume kinetic temperatures lower than 100~K. 

Equation~(\ref{f.c+cf+2}) implies that the $\frac{n({\rm CF}^{+})}{n({\rm C}^{+})}$ ratio is directly proportional to the chosen fluorine abundance. 
If we consider a fluorine abundance of $\lesssim 2.2\times 10^{-9}$ in molecular clouds (see Sect.~\ref{sec.fluorine}) instead of the fluorine abundance measured toward red giants \citep{2003AJ....126.1305C}, the estimated equilibrium ratio decreases to $\lesssim 1\times 10^{-6}$, which is comparable to the observed ratio. Therefore, we suggest that the low CF$^{+}$/C$^{+}$ column density ratios are expected in dense PDRs where CF$^{+}$ deficiency is observed. 






\section{Summary and conclusion}\label{Sec:sum}
In this work, we conducted targeted observations with the APEX 12~m sub-mm telescope towards N113, a region of active star-formation, located in the Large Magellanic Cloud (LMC). Our observations have led to the successful detection of the $J=2\to 1$ and $3\to 2$ lines of CF$^{+}$, representing the first discovery of this molecule in the LMC and in fact the second extragalactic detection. Our LTE analysis indicates an excitation temperature of 11.1--13.6~K and a CF$^{+}$ column density of $(0.6-2.4)\times 10^{12}$~cm$^{-2}$ for an assumed source size of 10\arcsec--23\rlap{.}\arcsec5. Based on non-LTE models, we infer that CF$^{+}$ emission emerges from dense PDRs with an H$_{2}$ number density of $(0.5-7.9)\times 10^{4}$~cm$^{-3}$ in N113. Based on a simple chemical network for CF$^{+}$, we further infer a fluorine abundance of $\lesssim 1.7\times 10^{-9}$, which is about an order of magnitude lower than measurements toward red giants in the LMC. Such CF$^{+}$ deficiency also give rise to a lower CF$^{+}$/C$^{+}$ column density ratio in dense PDRs. Therefore, we propose that CF$^{+}$ measurements can exclusively serve as conservative estimates, establishing lower bounds for both fluorine abundances and C$^{+}$ column densities in galaxies. 
Our study highlights the utility of CF$^{+}$ transitions, even when offering only lower limits, as an alternative tracer for probing the abundances of ionized carbon and fluorine. This opens a new avenue to assess chemical evolution in low-metallicity galaxies in the early Universe.


\section*{ACKNOWLEDGMENTS}\label{sec.ack}
We express our gratitude to the APEX staff for their invaluable assistance throughout our observations. We thank Prof. David A. Neufeld for his insightful comments on this draft. C.-H.R. C. acknowledges support from the Deutsches Zentrum für Luft- und Raumfahrt (DLR) grant NS1 under contract no. 50 OR 2214. The data was collected under the Atacama Pathfinder EXperiment (APEX) project, led by the Max Planck Institute for Radio Astronomy under the umbrella of the ESO La Silla Paranal Observatory. This research has made use of NASA's Astrophysics Data System. This work also made use of Python libraries including Astropy\footnote{\url{https://www.astropy.org/}} \citep{2013A&A...558A..33A}, NumPy\footnote{\url{https://www.numpy.org/}} \citep{5725236}, SciPy\footnote{\url{https://www.scipy.org/}} \citep{jones2001scipy}, and  Matplotlib\footnote{\url{https://matplotlib.org/}} \citep{Hunter:2007}. We would like to thank the referee for the valuable comments that have contributed to the improvement of this paper.

\bibliographystyle{aa}
\bibliography{references}

\begin{appendix}
\section{Hydrogen column density}\label{app.a}
To determine the hydrogen column density toward our N113 position, we employed a modified-blackbody emission model \citep[e.g.,][]{2008A&A...487..993K,2017ApJ...840...22L}:  
\begin{equation}\label{e.f1}
S_{\nu} = \Omega B_{\nu}(T_{\rm d})(1-e^{-\tau_{\nu}})\;,
\end{equation}
where $\Omega$ represents the solid angle, $B_{\nu}(T_{\rm d})$ is the Planck function at a dust temperature of $T_{\rm d}$, and $\tau_{\nu}$ is the optical depth of the dust emission at the observed frequency, $\nu$; the total hydrogen column density, $N_{\rm tot}({\rm H})$, is expressed as: 
\begin{equation}\label{e.f2}
N_{\rm tot}({\rm H}) = \frac{r_{\rm GDR}\tau_{\nu}}{\kappa_{\nu} m_{\rm H}}\;,
\end{equation}
where the gas-to-dust ratio, $r_{\rm GDR}$, is assumed to be 300 in the LMC \citep{2014ApJ...797...86R,2022ApJ...931..102S},  $\kappa_{\nu}$ is the dust opacity at the given frequency, and $m_{\rm H}$ represents the mass of a hydrogen atom. The dust opacity is modeled as $\kappa_{\rm \nu} = \kappa_{230}(\frac{\nu}{\rm 230})^{\beta}$ \citep{1983QJRAS..24..267H}, where $\kappa_{230}$ is 0.8~cm$^{2}$~g$^{-1}$ at 230~GHz according to \citet{1994A&A...291..943O}, and $\beta$ is the dust spectral index.

\begin{figure}[!htbp]
\centering
\includegraphics[width = 0.49 \textwidth]{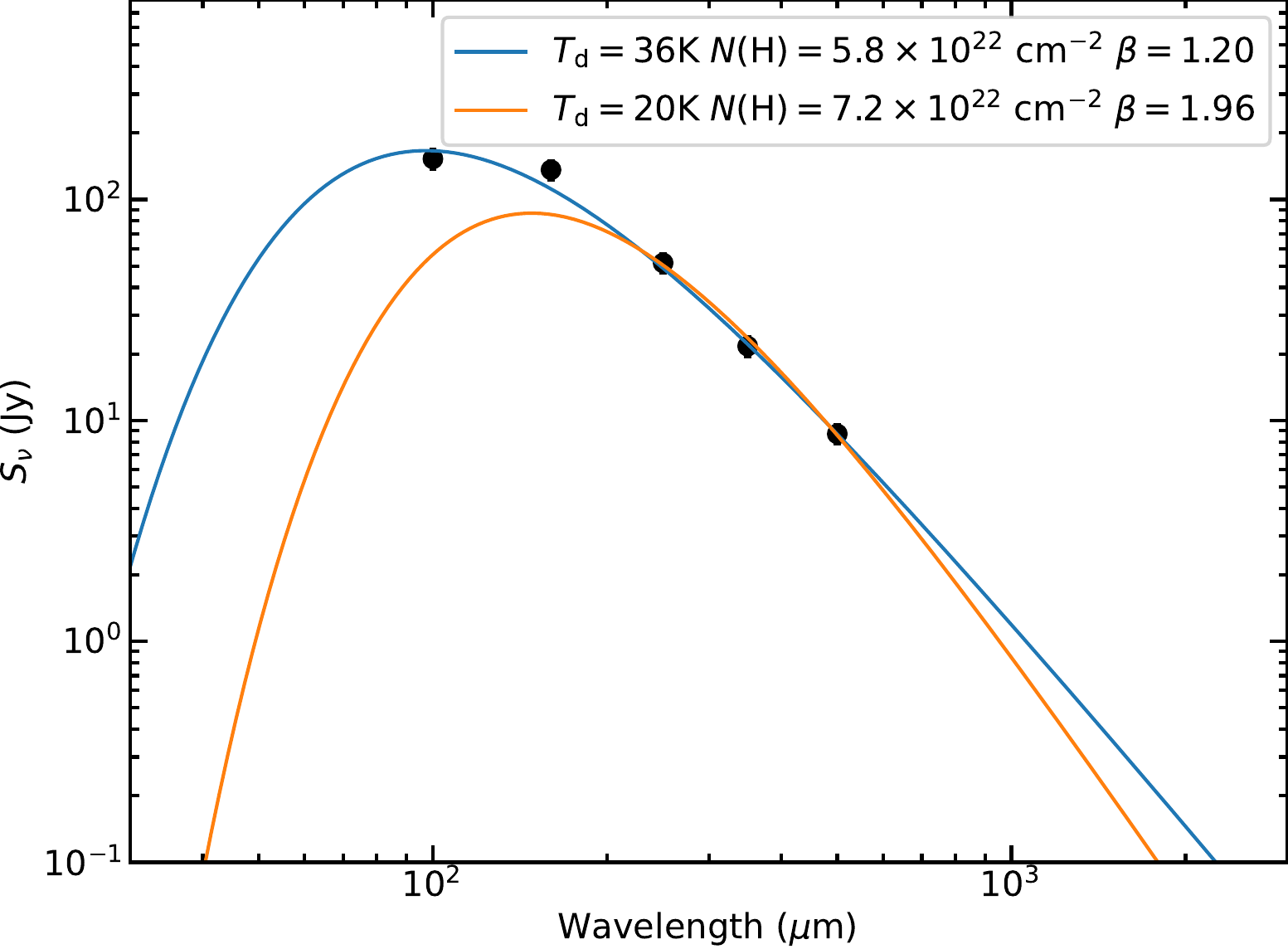}
\caption{{Spectral energy distribution of N113 fitted with a single modified blackbody model. The fitted values are indicated in the upper right corner.}\label{Fig:sed}}
\end{figure}

In this study, we utilized \textit{Herschel} continuum data\footnote{\url{https://karllark.github.io/data_magclouds_dustmaps.html}} \citep{2014ApJ...797...85G}, with foreground emission from Milky Way dust subtracted. This data set has a HPBW of 40\arcsec, corresponding to a solid angle of 4.26$\times 10^{-8}$~sr. We used two approaches for the spectral energy distribution (SED) fitting of N113. We first fixed the dust spectral index to be 1.96 for the fitting, and the result is indicated by the orange curve in Fig.~\ref{Fig:sed}. The resulting parameters are $T_{\rm d}=19.8\pm2.2$~K and $N_{\rm tot}({\rm H})=(7.2\pm 1.6)\times 10^{22}$~cm$^{-2}$. However, this fit does not match the 100~$\mu$m and 160~$\mu$m data points well, which could be the so-called submillimeter flattening probably caused by different dust components \citep[e.g.,][]{2019A&A...627A..15P}. Hence, the fitting procedure was repeated with the dust spectral index considered as a free parameter, yielding the blue curve in Fig.~\ref{Fig:sed}. This approach resulted in $T_{\rm d}=36.0\pm 5.4$~K, $N_{\rm tot}({\rm H})=(5.8\pm 0.6)\times 10^{22}$~cm$^{-2}$, and $\beta=1.18\pm 0.15$. Despite variations in $T_{\rm d}$ and $\beta$, the hydrogen column densities obtained through the two approaches are found to be comparable. Considering that the latter approach matches the data better, we chose $N_{\rm tot}({\rm H})=N({\rm H})+2N({\rm H_{2}})=(5.8\pm 0.6)\times 10^{22}$~cm$^{-2}$ for our analysis, where $N({\rm H})$ is the hydrogen column density in the atomic phase and $N({\rm H_{2}})$ is the H$_{2}$ column density (i.e., the hydrogen column density in the molecular phase). Because the hydrogen column density in the atomic phase is $N({\rm H})\sim 2.6\times 10^{21}$~cm$^{-2}$ from N113 \citep{2003ApJS..148..473K}, the hydrogen column density in the molecular phase is thus $N({\rm H_{2}})\sim 2.8\times 10^{22}$~cm$^{-2}$ and the corresponding $N({\rm H})/N({\rm H_{2}})$ ratio is $\sim$10\%. This ratio manifests the dominance of the molecular phase in N113.



\end{appendix}

\end{document}